\documentclass{webofc}
\usepackage{xspace}
\newcommand{\gev}{\,\mathrm{GeV}} 
\newcommand{\mev}{\,\mathrm{MeV}} 
\newcommand{\compass}{\textsc{Compass}\xspace}

\usepackage[varg]{txfonts}   
\woctitle{New Frontiers in Physics 2014}
\begin{document}
\title{Hadron Physics at the \textsc{Compass} Experiment
}
%
%

\author{Fabian Krinner for the \textsc{Compass} collaboration\inst{1}\fnsep\thanks{\email{fabian-krinner@mytum.de}}
}

\institute{Technische Universit\"at M\"unchen, Physik-Department, E18
          }

\abstract{%
Quantum Chromodynamics (QCD), the theory of strong interactions, in
principle describes the interaction of quark and gluon fields. However,
due to the self-coupling of the gluons, quarks and gluons are confined
into hadrons and cannot exist as free particles. The quantitative
understanding of this confinement phenomenon, which is responsible for
about 98\% of the mass of the visible universe, is one of the major open
questions in particle physics. The measurement of the excitation
spectrum of hadrons and of their properties gives valuable input to
theory and phenomenology.

In the Constituent Quark Model (CQM) two types of hadrons exist: mesons,
made out of a quark and an antiquark, and baryons, which consist of
three quarks. But more advanced QCD-inspired models and Lattice QCD
calculations predict the existence of hadrons with exotic properties
interpreted as excited glue (hybrids) or even pure gluonic bound states
(glueballs).

The \textsc{Compass} experiment at the CERN Super Proton Synchrotron has acquired
large data sets, which allow to study light-quark meson and baryon
spectra in unprecedented detail. The presented overview of the first
results from this data set focuses in particular on the light meson sector 
and presents a detailed analysis of three-pion final states. A new $J^{PC} = 1^{++}$ state, the $a_1(1420)$, is observed with a mass and width in the ranges $m = 1412-1422\mev/c^2$ 
and $\Gamma = 130-150\mev/c^2$.
}
\maketitle
\section{The \textsc{Compass} experiment}
\label{sec::compass}

The \compass experiment is a multi-purpose fixed-target spectrometer located at CERN's Prevessin-area. It is supplied with secondary hadron and tertiary muon beams by the Super Proton Synchrotron. \compass is a two-stage spectrometer that covers a wide kinematic range and employs beam and final-state particle identification using Cherenkov-detectors (CEDARs and RICH, respectively). The wide physics program includes studies of the nucleon spin-structure, as well as hadron spectroscopy. The latter is presented here.\\
The presented analysis is performed on data taken in 2008, where a $190\gev/c$ negative secondary hadron beam, consisting mainly of $\pi^-$ ($97\%$) with some admixture of negative kaons ($2\%$) and antiprotons ($1\%$), impinged on a $40\,\mathrm{cm}$ liquid-hydrogen target.\\

The analysis described here was performed on two different three pion final-states, namely $\pi^-p\to \pi^-\pi^0\pi^0p$ and $\pi^-p\to\pi^-\pi^+\pi^-p$. For these final states, $3.5$ million events were recorded in the neutral channel, $\pi^-\pi^0\pi^0$, and $50$ million in the charged channel, $\pi^-\pi^+\pi^-$. The latter constitutes at the moment the world's largest $3\pi$ data set. 
\\The anlyses of both channels were performed independently, using different software packages. Since the reconstruction of the two channels also relies on different parts of the spectrometer, the systematic uncertainties differ.
Nevertheless, since both final states only vary in their decay modes, both data sets are expected to contain the same resonant structures.
\begin{figure}[bt]
\centering
\begin{minipage}{0.39\textwidth}
\includegraphics[width=\linewidth]{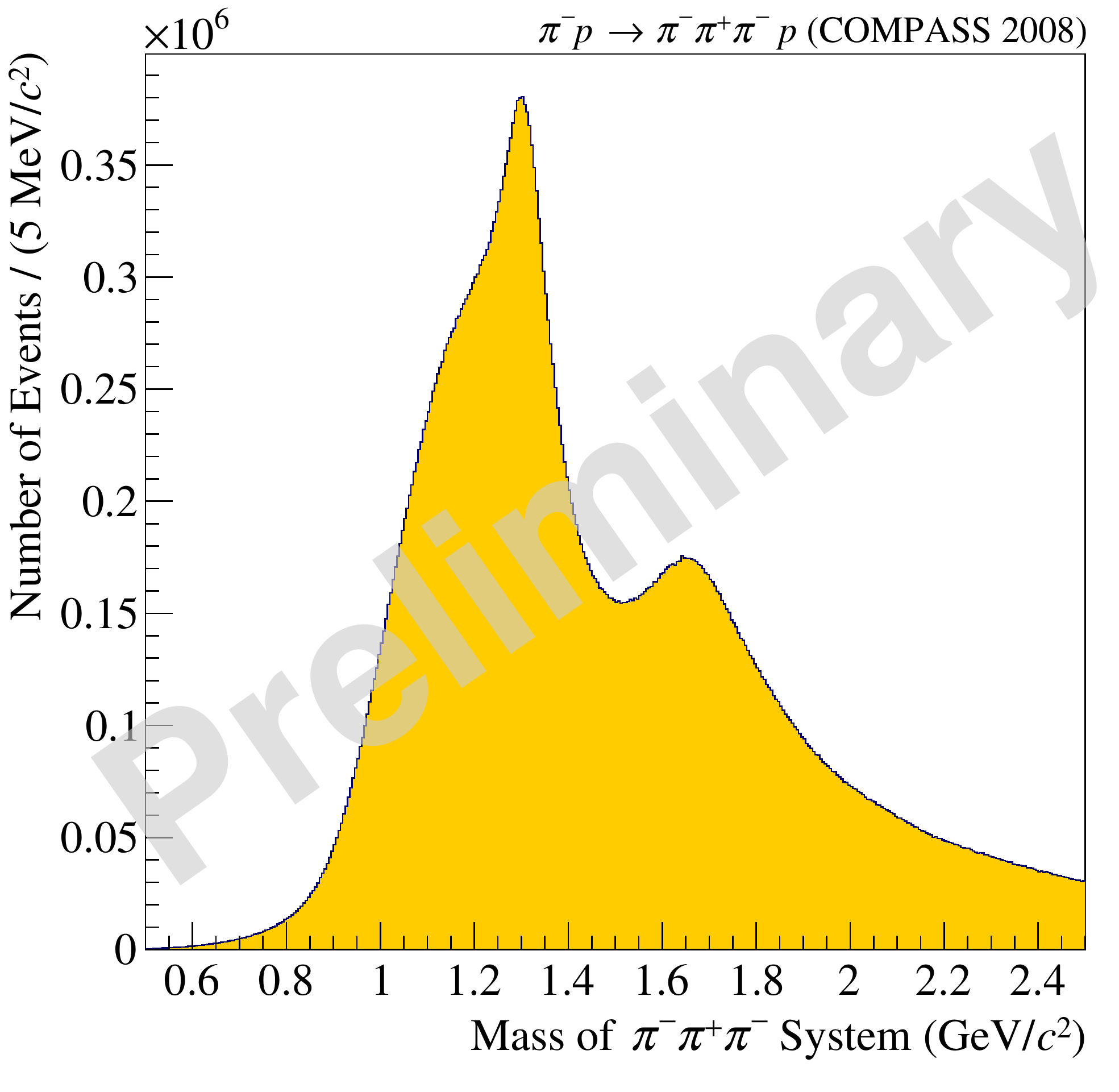}
\caption{Mass spectrum of the $\pi^-\pi^+\pi^-$ final state.\vspace*{2\baselineskip}}
\label{pic::specc}
\end{minipage}
\begin{minipage}{0.05\textwidth}
 \hspace{0.05\textwidth}
\end{minipage}
\begin{minipage}{0.49\textwidth}
\centering
\includegraphics[width=\linewidth]{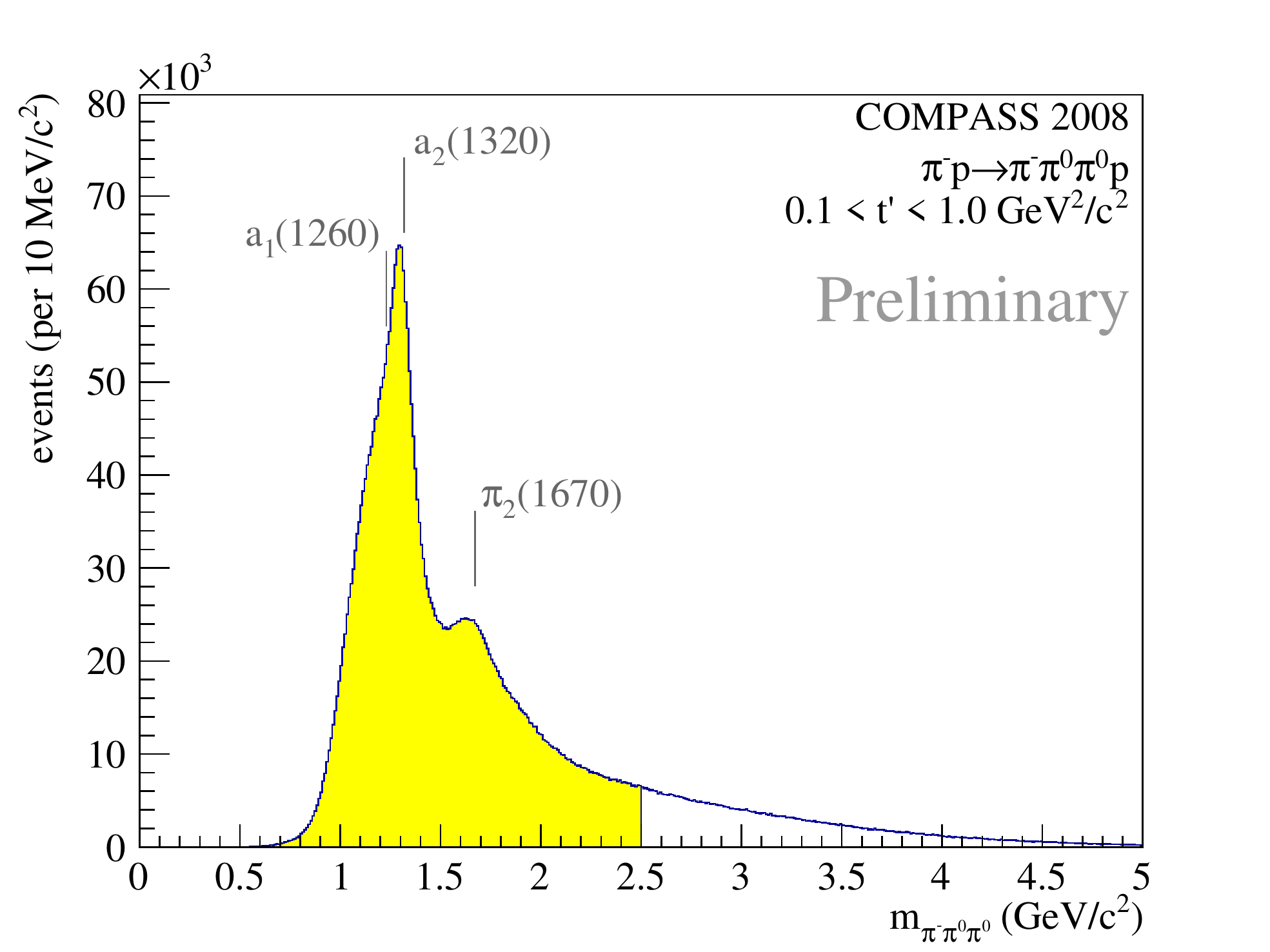}
\caption{Mass spectrum of the $\pi^-\pi^0\pi^0$ final state. The main structures are labeled by the corresponding resonances described in sec. \ref{sec::major}.}
\label{pic::specn}
\end{minipage}
\end{figure}

\section{Partial-wave analysis}
\label{sec::pwa}
For the following studies, a Partial-Wave Analysis (PWA) was performed on diffractively produced three-pion final states. In this process, the incoming beam pion gets excited via Pomeron exchange with the target proton into an intermediate state $X^-$\!\!, which then decays into the final state.\\
The intermediate state $X^-$ is characterized by the quantum numbers $J^{PC}M^\epsilon$ of $X^-$\!\!, where $J$ is the spin, $P$ and $C$ the eigenvalues of parity and generalized charge conjugation, $M$ the magnetic quantum number and $\epsilon$ the reflectivity of $X^-$\!\!.\\
These quantum numbers are not known a priori and there are many different values they can take. Since inital and final state are always identical all these different possibilities may interfere with each other.\\
The main goal of the analysis is to disentangle the contributions of all these intermediate states, which is achieved via PWA.\\
\begin{figure}[bt]
\centering
\begin{minipage}{0.44\textwidth}
\includegraphics[width=\linewidth]{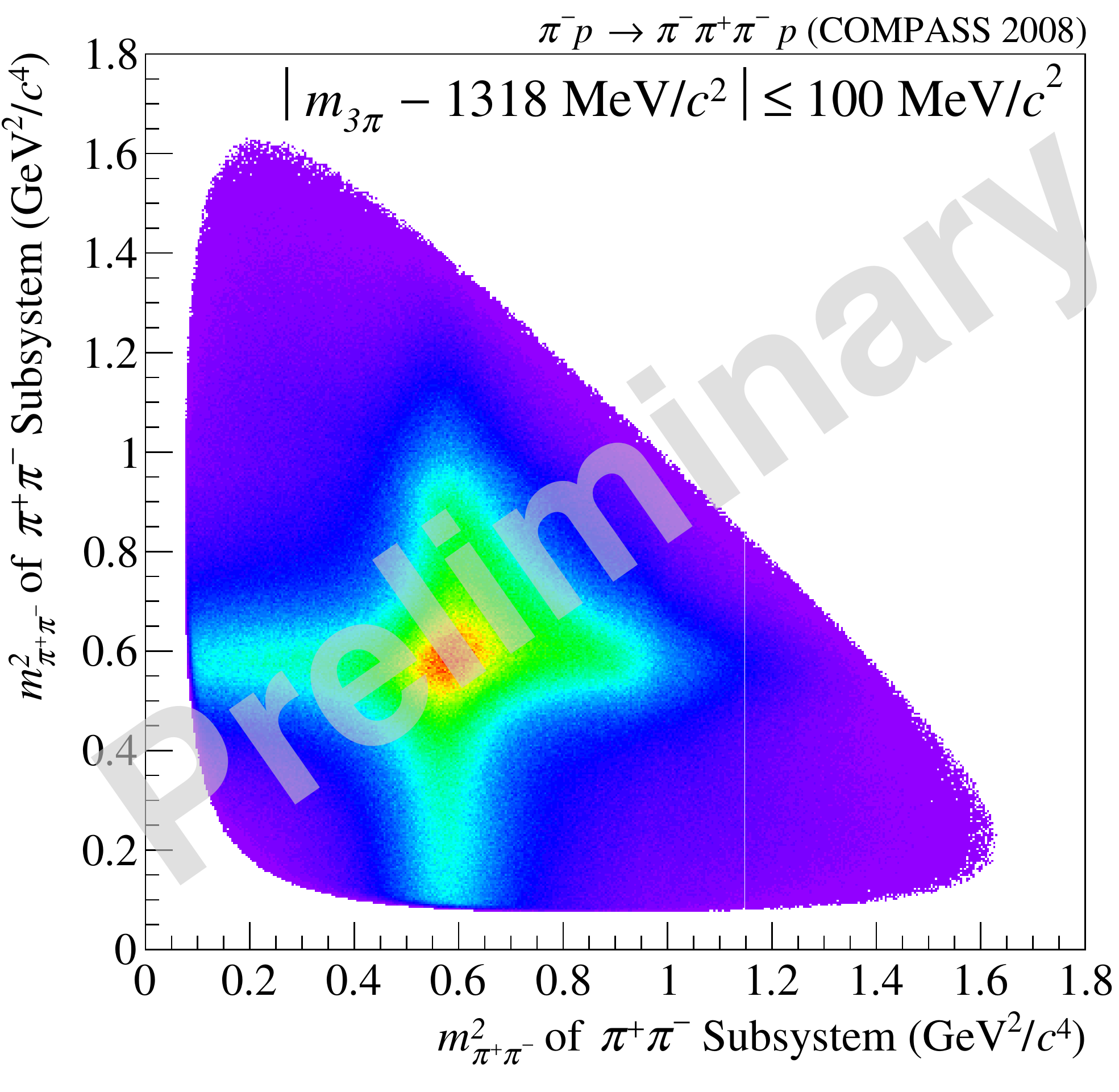}
\caption{Dalitz plot of the $\pi^-\pi^+\pi^-$ channel, with $m_{3\pi}$ around the mass of the $a_2(1320)$}
\label{pic::specc}
\end{minipage}
\begin{minipage}{0.05\textwidth}
 \hspace{0.05\textwidth}
\end{minipage}
\begin{minipage}{0.44\textwidth}
\centering
\includegraphics[width=\linewidth]{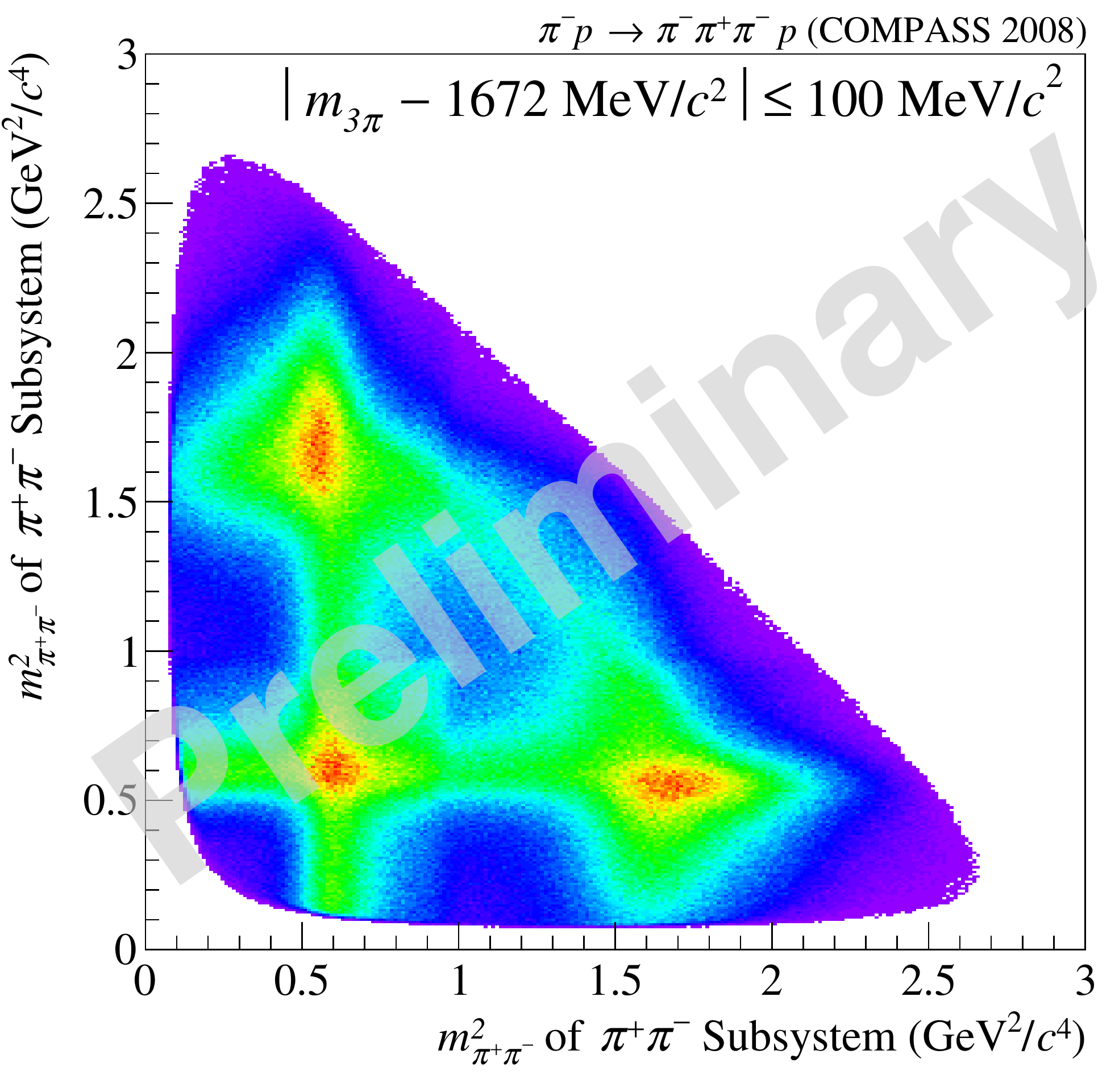}
\caption{Dalitz plot of the $\pi^-\pi^+\pi^-$ channel, with $m_{3\pi}$ around the mass of the $\pi_2(1670)$}
\label{pic::specn}
\end{minipage}
\end{figure}
\subsection{The {isobar}-model}
\label{sec::isobar}
For the PWA, the \textit{isobar} model is employed. In this model, it is assumed that the diffractively produced state $X^-$ does not decay directly into the multi-particle final state, but undergoes subsequent two-particle decays until it reaches the final state. In the case of three pions, $X^-$ decays first into a bachelor $\pi$ and another intermediate state $\xi$, the so called {isobar}, which then decays into two pions.\\
Since in the {isobar} model, production and decay of $X^-$ factorize, the complex amplitude $\mathcal{A}$ of the process can be expanded into a series of partial waves:
\begin{equation}\label{eq::amp}
\mathcal{A}(m_X,\tau) = \sum_\mathrm{waves} T_\mathrm{wave}(m_X) \psi_\mathrm{wave}(\tau)
,\end{equation}
where the decay amplitudes $\psi_\mathrm{wave}(\tau)$, which depend on the phase-space variables $\tau$, describe the kinematic distribution of the final-state particles. The production amplitudes $T_\mathrm{wave}$ describe the production of intermediate states $X^-$ with certain quantum numbers $J^{PC}M^\epsilon$\!\!. With these amplitudes, the observed intensity $\mathcal{I}$ of the process is given as:
\begin{equation}\label{eq::intens}
\mathcal{I}(m_X,\tau) = \left|\mathcal{A}(m_X,\tau)\right|^2 = \Bigg|\sum_\mathrm{waves} T_\mathrm{wave}(m_X) \psi_\mathrm{wave}(\tau)\Bigg|^2
.\end{equation}
In the framework of the isobar model, the decay amplitudes $\psi(\tau)$ are calculable, if one puts in fixed parametrizations for the line shapes of the isobars. For the current analysis, the following six {isobars} were used \cite{Au:1986vs,Beringer:1900zz,Flo:Thesis}:
\begin{center}
 \begin{tabular}{ c|l }
  $I^GJ^{PC}$ &   \\
  \hline
  $ 0^+0^{++}$ & $[\pi\pi]_S,\,f_0(980),\,f_0(1500)$  \\
  $ 1^+1^{--}$ & $\rho(770)$  \\
  $ 0^+2^{++}$ & $f_2(1270)$  \\
  $ 1^+3^{--}$ & $\rho_3(1690)$ \\
 \end{tabular}
\end{center}
With known $\psi(\tau)$, the production amplitudes $T(m_X)$ can be determined by fitting eq. (\ref{eq::intens}) to the observed intensity distribution in bins of $m_{3\pi} = m_X$. Therefore, no assumptions about the $3\pi$ resonances have to be made.
\subsection{The wave set}
The amplitude described in equation (\ref{eq::amp}) employs a sum over different $\mathrm{waves}$. These waves are defined by:
\begin{equation}
J^{PC}M^\epsilon [\mathrm{isobar}]\ \pi\ L
,\end{equation}
where $J^{PC}M^\epsilon$ are the quantum numbers of $X^-$\!\!. Since the {isobars} are well-known states, their quantum numbers are known and thus not explicitly stated in the formula above. Finally, $L$ is the relative orbital angular momentum between the {isobar} and the bachelor pion.\\
For the current analysis a set of $87$ such waves with spin $J$ and angular momentum $L$ up to six was used, employing the {isobars} listed above. In addition, one incoherent isotropic wave was added in order to describe uncorrelated events. \cite{Flo:Thesis}
\label{sec::waveset}

\section{Results for single waves}
\label{sec::results}
\subsection{The major waves}
\label{sec::major}
\begin{figure}[bt]
\centering
\begin{minipage}{0.44\textwidth}
\includegraphics[width=\linewidth]{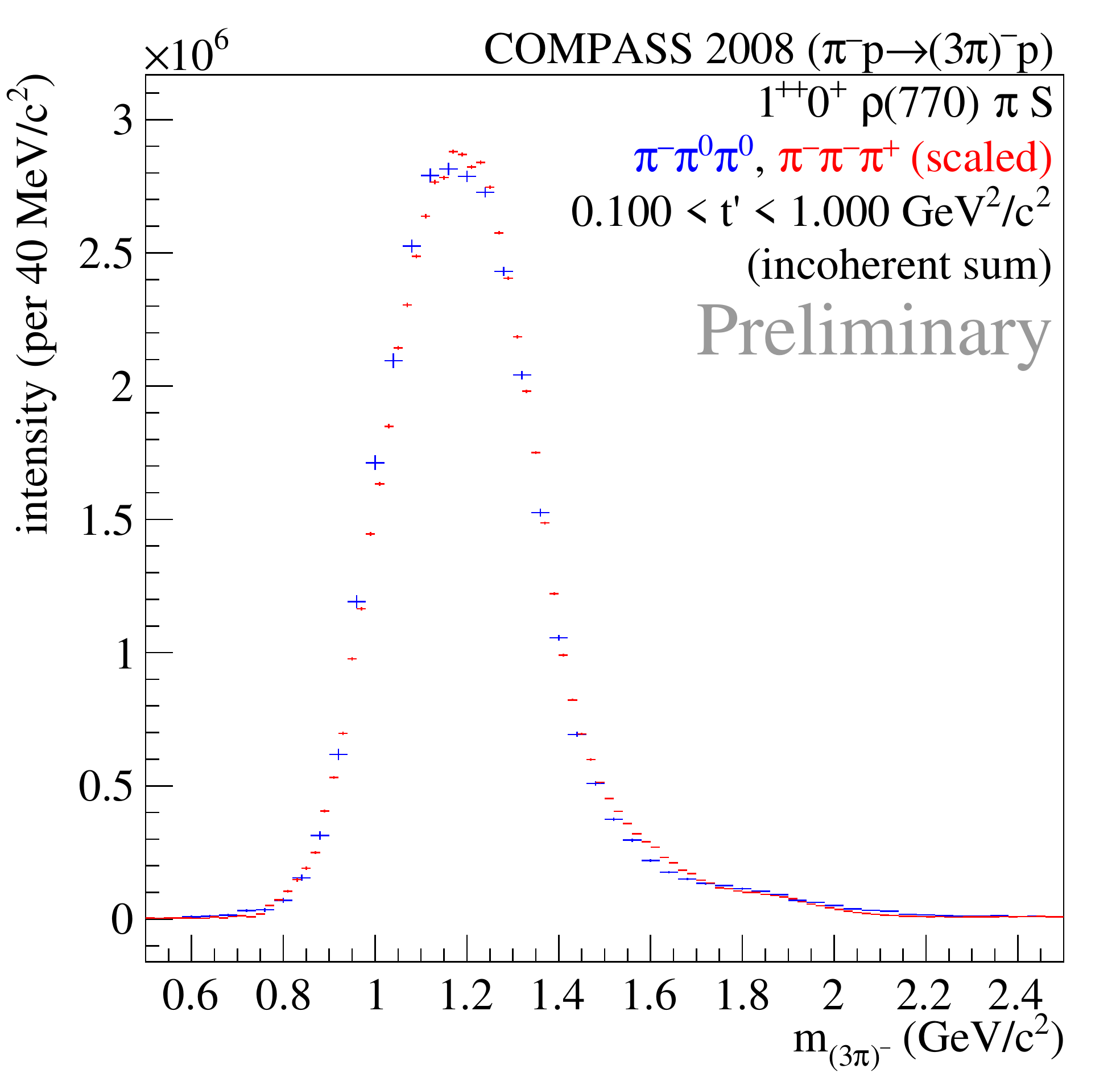}
\caption{Intensity of the $1^{++}0^+\rho(770)\ \pi\ S$ wave for both three-pion channels. The charged channel is scaled to match the intensity integral of the neutral one in this wave.}
\label{pic::1pp}
\end{minipage}
\begin{minipage}{0.05\textwidth}
 \hspace{0.05\textwidth}
\end{minipage}
\begin{minipage}{0.44\textwidth}
\centering
\includegraphics[width=\linewidth]{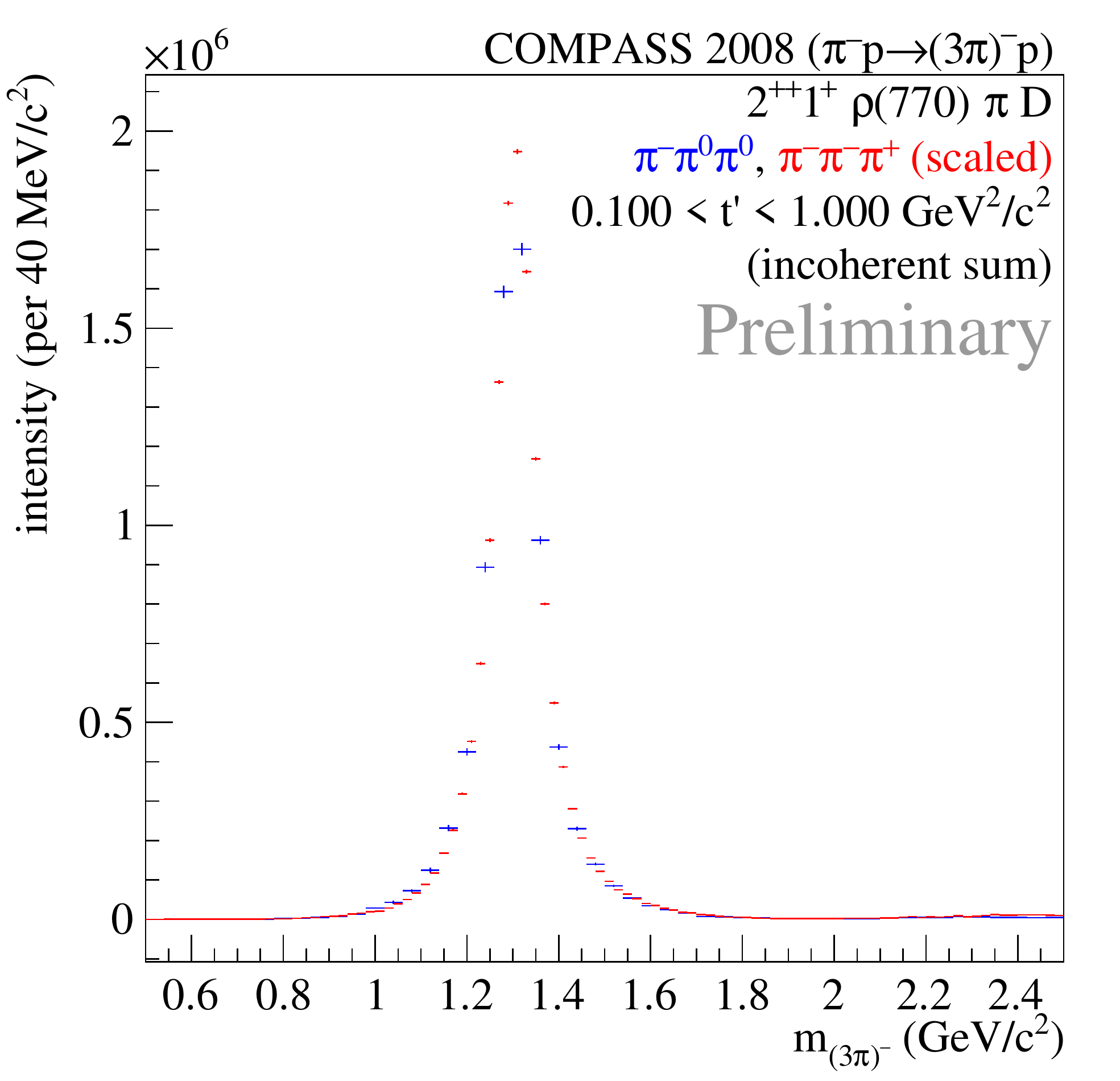}
\caption{Intensity of the $2^{++}1^+\rho(770)\ \pi\ D$ wave for both three-pion channels. The charged channel is scaled to match the intensity integral of the neutral one in this wave.}
\label{pic::2pp}
\end{minipage}
\end{figure}
\begin{figure}[bt]
\centering
\begin{minipage}{0.44\textwidth}
\includegraphics[width=\linewidth]{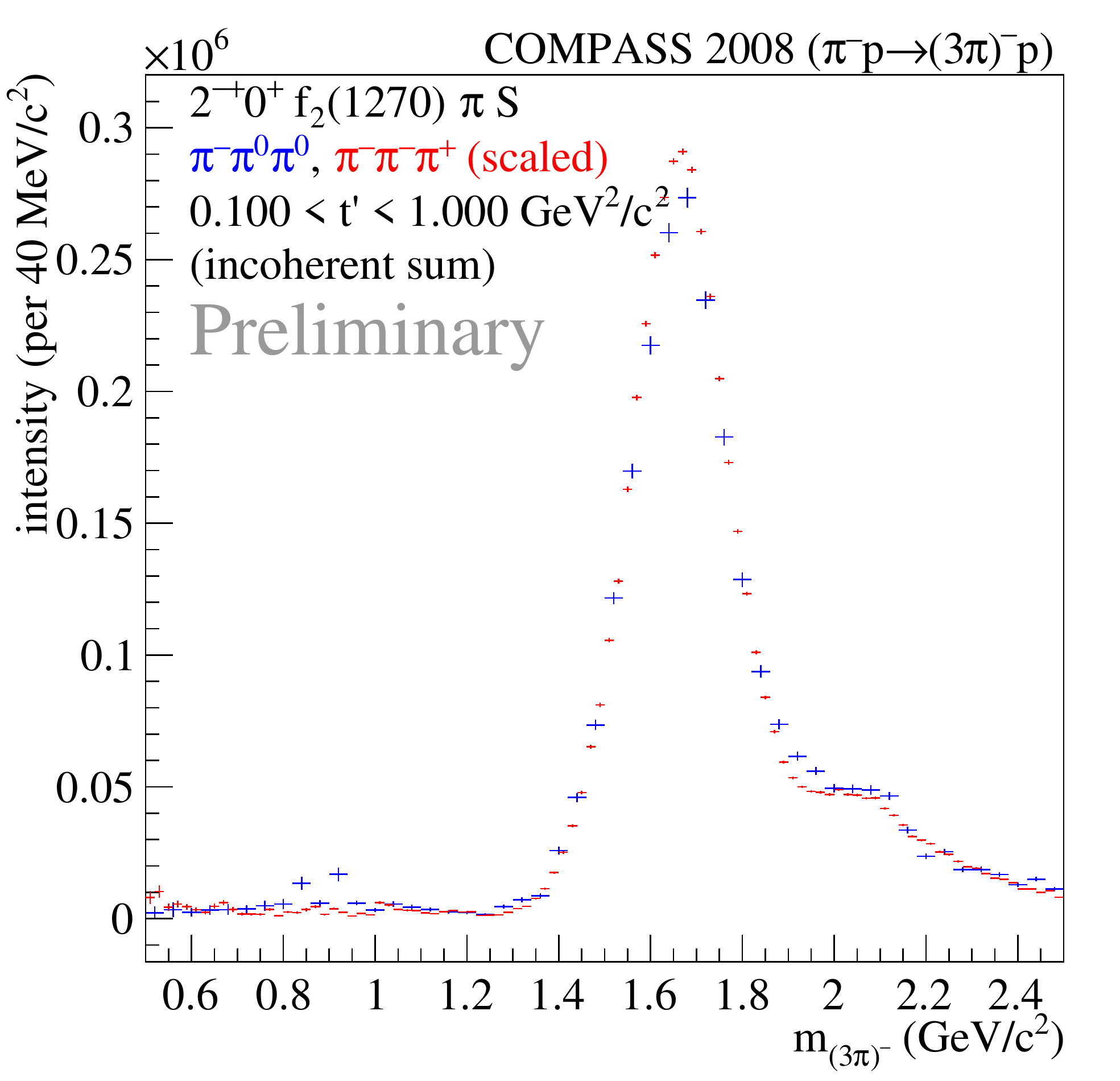}
\caption{Intensity of the $2^{-+}0^+f_2(1270)\ \pi\ S$ wave for both three-pion channels. The charged channel is scaled to match the intensity integral of the neutral one in this wave.}
\label{pic::2mp}
\end{minipage}
\begin{minipage}{0.05\textwidth}
 \hspace{0.05\textwidth}
\end{minipage}
\begin{minipage}{0.44\textwidth}
\centering
\includegraphics[width=\linewidth]{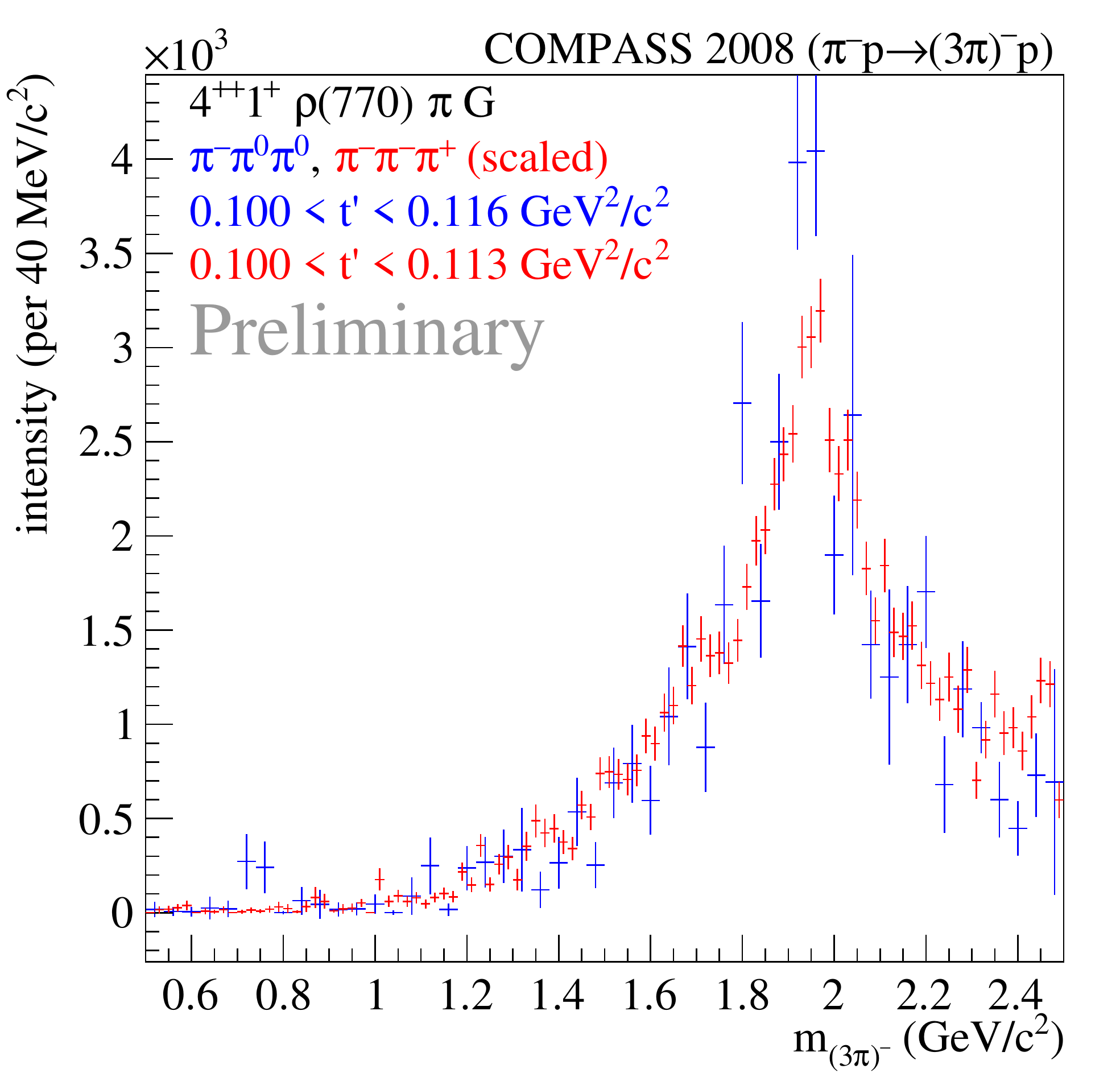}
\caption{Intensity of the $4^{++}1^+\rho(770)\ \pi\ G$ wave for both three-pion channels. The charged channel is scaled to match the intensity integral of the neutral one in this wave.}
\label{pic::4pp}
\end{minipage}
\end{figure}
The $3\pi$ invariant mass spectra for both channels, as shown in Figs. \ref{pic::specc} and \ref{pic::specn}, already exhibit some structures. They are explained by the three biggest waves in the analysis, namely:
\begin{eqnarray}
1^{++}0^+&\!\!\!\!\!\!\!\rho(770)\ \pi\ S\nonumber\\
2^{++}1^+&\!\!\!\!\!\!\!\rho(770)\ \pi\ D\nonumber\\
2^{-+}0^+&\!\!\!\!\!\!f_2(1270)\ \pi\ S\nonumber
\end{eqnarray}
The intensity of the first wave, $1^{++}0^+\rho(770)\ \pi\ S$, is depicted in Fig. \ref{pic::1pp}. The wave describes an axial-vector intermediate state decaying into $\rho(770)$ and a pion. It is the biggest wave in the analysis and takes about $33\%$ of the total intensity in the charged channel. The biggest structure visible in this wave is the $a_1(1260)$ resonance. The results for the neutral and charged channel are in good agreement.\\
The second biggest wave, $2^{++}1^+\rho(770)\ \pi\ D$, describes a spin-2 meson also decaying into $\rho(770)$ and $\pi$. Its intensity is depicted in Fig. \ref{pic::2pp}. This wave shows the clearest $3\pi$ resonance, the well-known $a_2(1320)$, with nearly no background. Again, neutral and charged channel agree well. This $2^{++}$ wave takes about $8\%$ of the total intensity in the charged channel.\\
The third largest wave is the $2^{-+}0^+f_2(1270)\ \pi\ S$, which takes approximately $7\%$ of the charged intensity. It describes a state with the quantum numbers of a pion with spin 2, decaying into $f_2(1270)\ \pi$. Its intensity is depicted in Fig. \ref{pic::2mp} which again shows good agreement in both channels. The main structure visible is the $\pi_2(1670)$.\\
The analysis performed is not only able to extract the major waves, but can also separate out small contributions on the sub-percent level. For example the $4^{++}1^+\rho(770)\ \pi\ G$ wave, describing a spin-4 state decaying into $\rho(770)\ \pi$, takes only about $0.76\%$ of the total intensity. Nevertheless, as can be seen in Fig. \ref{pic::4pp}, the $a_4(2040)$ is clearly visible in both channels.
\subsection{The $\mathbf{a_1(1420)}$}
\begin{figure}[bt]
\centering
\begin{minipage}{0.44\textwidth}
\includegraphics[width=\linewidth]{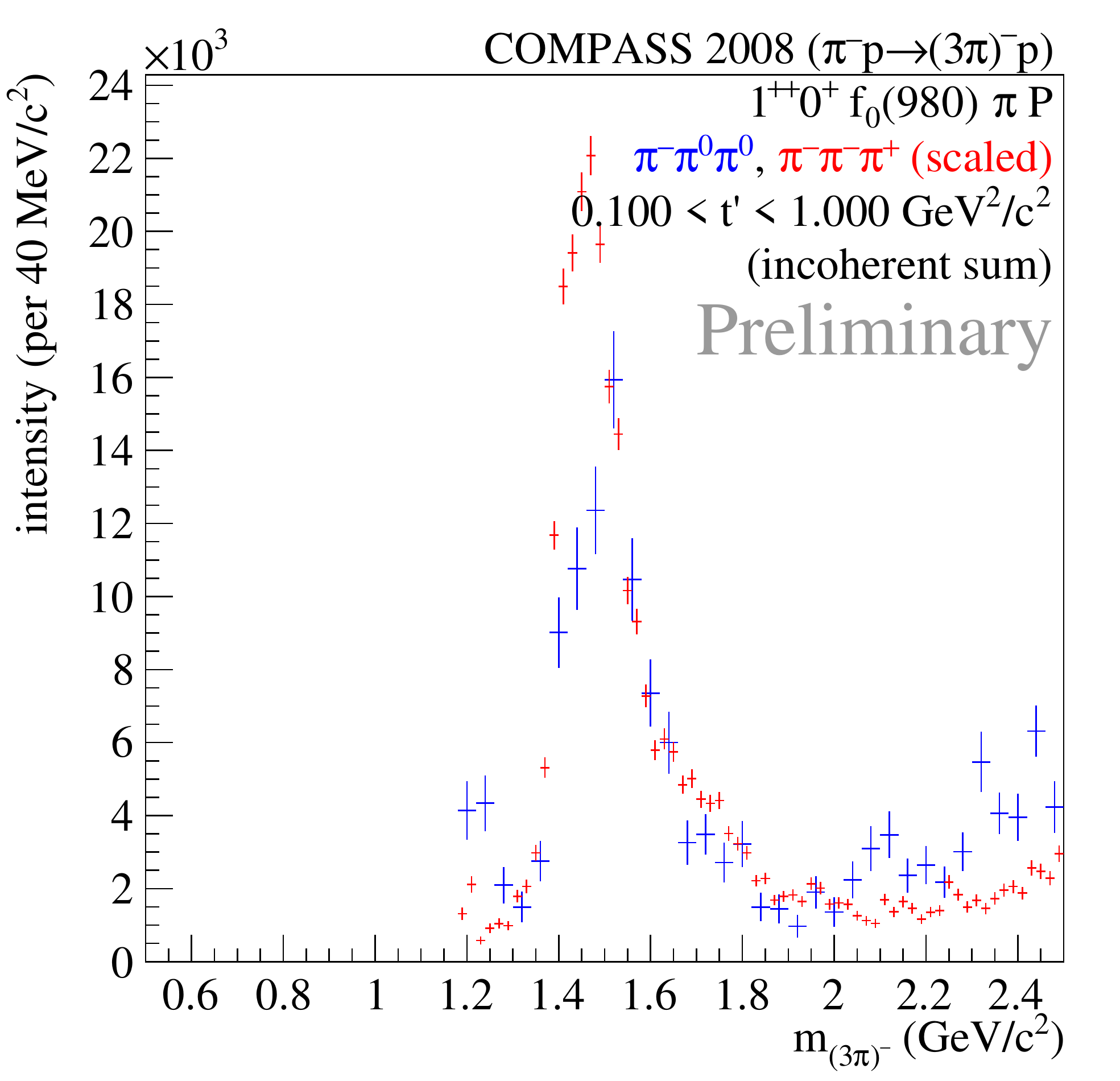}
\caption{Intensity of the $1^{++}0^+f_0(980)\ \pi\ P$ wave. The peak of the $a_1(1420)$ is clearly visible in both channels.\vspace*{\baselineskip}}
\label{pic::1420}
\end{minipage}
\begin{minipage}{0.05\textwidth}
 \hspace{0.05\textwidth}
\end{minipage}
\begin{minipage}{0.44\textwidth}
\centering
\includegraphics[width=\linewidth]{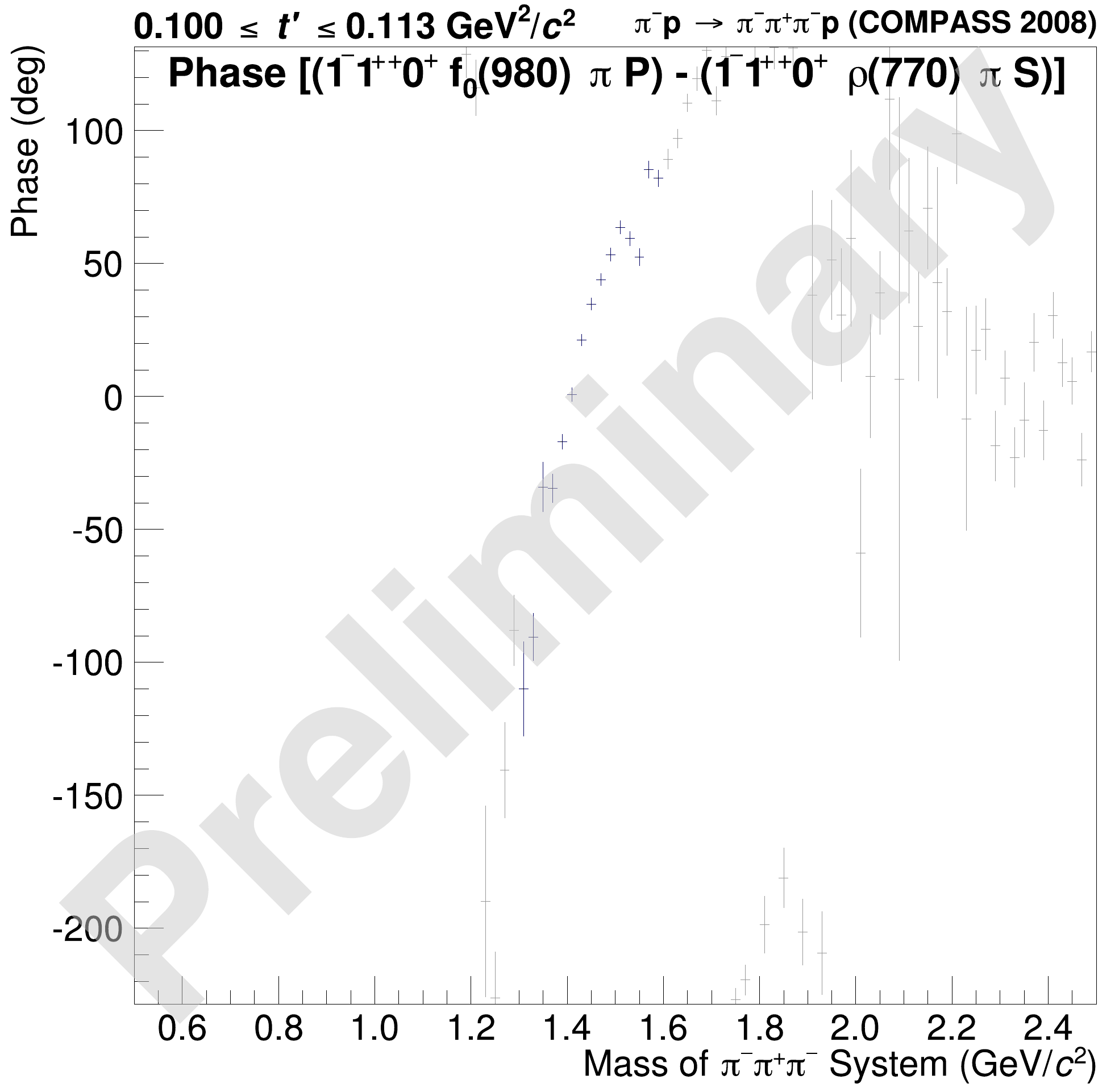}
\caption{Relative phase motion between the $1^{++}0^+\rho(770)\ \pi\ S $ and the $1^{++}0^+f_0(980)\ \pi\ P$ wave. In the region of the $a_1(1420)$, a rapid phase motion is visible.}
\label{pic::1260:1420}
\end{minipage}
\end{figure}
Besides the known resonances shown in the waves above, a previously unknown resonance is seen in the $1^{++}0^+f_0(980)\ \pi\ P$ wave. The state $X^-$ in this wave has the same quantum numbers ($J^{PC}M^\epsilon =  1^{++}0^+$) as in the biggest wave, which is dominated by the $a_1(1260)$, but differs in the decay channel, which is $f_0(980)\ \pi$ instead of $\rho(770)\pi$.\\
The $a_1(1420)$ can be clearly seen in both final-states, as shown in Fig. \ref{pic::1420}. It shows a clear and rapid phase motion with respect to the $1^{++}0^+\rho(770)\ \pi\ S$ wave, depicted in Fig. \ref{pic::1260:1420}, which indicates, that the observed structure is indeed a resonance. The Breit-Wigner resonance parameters were determined to be in the following ranges:
\begin{eqnarray}
m &=& 1412-1422\mev/c^2\\
\Gamma &=& 130-150\mev/c^2
\end{eqnarray}
The observation of this resonance in this mass region is peculiar for a number of reasons. First of all, no resonance has been predicted by models or lattice QCD in the $1.4\gev/c^2$ mass region. Second, this new resonance decays into $f_0(980)\pi$ with an unusually small intensity and the $f_0(980)$ is known to also couple strongly to $KK$. Last, the mass of the $a_1(1420)$ lies only sightly above the $KK^*$ threshold, a behavior which can also be seen in the $XYZ$ resonances in $D$ and $B$ physics.\\
Nevertheless, the nature of the $a_1(1420)$ is unclear at the moment. \cite{Basdevant:1977ya}
\label{sec::a1}

\section{Conclusions}
\label{sec::conclusion}
The \compass experiment has collected large data sets for the two three-pion channels $\pi^-\pi^0\pi^0$ and $\pi^-\pi^+\pi^-$, $3.5$ and $50$ million events respectively. For the charged channel, this is at the moment the world's largest data set. These data allow to perform a very detailed partial-wave analysis with a systematic cross check between the two channels and thus a deep insight into the light hadron spectrum.
In the present analysis, a wave-set of $87$ waves up to spin 6 was employed, and waves on the sub-percent level could be analyzed.
Besides showing clear signals of known states, a previously unknown resonance, the $a_1(1420)$, was seen in the $1^{++}0^+f_0(980)\ \pi\ P$ wave.

\section{Outlook}
\label{sec::outlook}
Based on the presented analysis, fits to the $m_{3\pi}$ dependencies of the intensities and relative phases of a subset of waves are performed, to disentangle resonant and non-resonant contributions and determine the parameters, i.e. masses and widths, of the observed resonances. \cite{Proceeding:Haas}\\
In addition to the waves shown above, a spin-exotic wave was part of the wave set containing a possible exotic resonance is studied. \cite{Proceeding:Haas}\\
Since the PWA relies on the {isobar}-model, further studies are being performed in order to determine the {isobar} line shapes from the data. If successful, this will allow us to check the validity of the {isobar} model and even extract resonance parameters of the {isobars}. \cite{Fabi:Bormio}\\


\begin{thebibliography}{99}

\bibitem{Beringer:1900zz}
  J.~Beringer {\it et al.}  [Particle Data Group Collaboration],
  Phys.\ Rev.\ D {\bf 86} (2012) 010001.

\bibitem{Au:1986vs}
  K.~L.~Au, D.~Morgan and M.~R.~Pennington,
  Phys.\ Rev.\ D {\bf 35} (1987) 1633.

\bibitem{Flo:Thesis}
 F.~Haas, {\it Two-Dimensional Partial-Wave Analysis
of Exclusive $190\gev$ $\pi^-p$ Scattering into
the $\pi^-\pi^-\pi^+$ Final State at \textsc{Compass} (CERN)}
, PhD Thesis, Technische Universit\"at M\"unchen, 2014.

\bibitem{Basdevant:1977ya}
  J.~L.~Basdevant and E.~L.~Berger,
  Phys.\ Rev.\ D {\bf 16} (1977) 657.


\bibitem{Proceeding:Haas} F.~Haas, Proceedings of the 13th International Workshop on Meson Production, Properties and Interaction (2014), Krak\'ow, Poland, to be published in {\it EPJ Web of Conferences}.

\bibitem{Fabi:Bormio}
 F.~Krinner,
 Proceedings of the 52nd International Winter Meeting on Nuclear Physics (2014), Bormio, Italy, to be published in {\it Proceedings of Science}.

\end{thebibliography}
\end{document}